\documentclass[a4paper,fleqn,usenatbib]{mnras}
\usepackage{newtxtext,newtxmath}

\usepackage[T1]{fontenc}
\usepackage{ae,aecompl}
\usepackage{color}

\usepackage{graphicx}	
\usepackage{amsmath}	
\usepackage{amssymb}	
\usepackage{subcaption}
\title[Full GR simulation of very massive star and the BH formation]{A full general relativistic neutrino radiation-hydrodynamics simulation of a collapsing very massive star and the
formation of a black hole}

\author[Kuroda, Kotake, Takiwaki \& Thielemann]{
Takami Kuroda,$^{1}$\thanks{E-mail: takami.kuroda@physik.tu-darmstadt.de}
Kei Kotake,$^{2}$
Tomoya Takiwaki$^{3}$ and
Friedrich-Karl Thielemann$^{4,5}$
\\
$^1$Institut f{\"u}r Kernphysik, Technische Universit{\"a}t Darmstadt, Schlossgartenstrasse 9, D-64289 Darmstadt, Germany\\
$^2$Department of Applied Physics, Fukuoka University, 8-19-1, Jonan, Nanakuma, Fukuoka, 814-0180, Japan\\
$^3$Division of Theoretical Astronomy, National Astronomical Observatory of Japan
 (NAOJ), 2-21-1, Osawa, Mitaka, Tokyo, 181-8588, Japan\\
$^4$Department of Physics, University of Basel, Klingelbergstrasse 82, CH-4056 Basel, Switzerland\\
$^5$GSI Helmholtzzentrum f\"ur Schwerionenforschung, Planckstrasse 1, D-64291 Darmstadt, Germany
}

\date{Accepted XXX. Received YYY; in original form ZZZ}

\pubyear{2017}

\begin{document}
\label{firstpage}
\pagerange{\pageref{firstpage}--\pageref{lastpage}}
\maketitle

\begin{abstract}
 We study the final fate of a very massive star by performing full general 
relativistic (GR), three-dimensional (3D) simulation with three-flavor
 multi-energy
neutrino transport. Utilizing a 70 solar mass zero metallicity progenitor,
 we self-consistently follow the radiation-hydrodynamics from 
the onset of gravitational core-collapse until the second collapse of 
the proto-neutron star (PNS), leading to black hole (BH) formation.
 Our results show that the BH formation occurs at a post-bounce time of 
$T_{\rm pb}\sim 300$ ms for the 70 $M_{\odot}$ star. This is significantly 
 earlier than those in the literature where lower mass progenitors were employed. 
  At a few $\sim10$ ms before BH formation, we find that the stalled 
 bounce shock is revived by intense neutrino heating from the very hot PNS, 
which is aided by violent convection behind the shock. 
In the context of 3D-GR core-collapse modeling with
 multi-energy neutrino transport, our numerical results present the first evidence to validate a {\it fallback} BH formation scenario of the $70 M_{\odot}$ star.
\end{abstract}




\section{Introduction}
\label{sec1}
It is now of great importance to unveil the origin of ``massive'' black holes
after the first gravitational-wave (GW) 
detection by the LIGO collaboration \citep{GW150914}.
 Since then, a rich variety of the BH masses ($\sim 8$ to $35 M_{\odot}$)
 have been discovered in binary black hole (BBH) merger events
 \citep[e.g.,][]{GW151226,GW170104}.

One of the most plausible scenarios to explain the BBHs 
 is a binary stellar evolution in a low-metallicity environment 
(see \citet{abbott16} for a review).
 It has been proposed that two massive stars 
in the approximate range of $40$ to $100 M_\odot$ lead to the formation 
of a massive helium core
(e.g., \citet{bel10,Langer12,Kinugawa16} for collective 
references therein).
The gravitational collapse of the massive core ($\sim30M_\odot$)
 could account for some of the relevant BH mass ranges (at least in 
 the high-mass end) in the GW events, although the formation path to
 the massive core and further on to the BH 
is still very uncertain due to the complexity of the binary 
evolution and the fallback dynamics (e.g., \citet{Fryer99,Heger03,bel14}).

In order to clarify the formation process of the BH,
one requires full general relativistic and neutrino 
radiation-hydrodynamics core-collapse simulations of such massive stars
 in three-dimensional space.
 Due to the high numerical cost 
(e.g., \citet{janka16,lentz15,Kotake12_ptep}),
 most of the previous studies with BH formation have been performed assuming
 spherical symmetry (1D) (e.g., 
\citet{Liebendorfer04,Sumiyoshi07,Fischer09,O'Connor11}).
In multi-dimensional (multi-D) simulations, \cite{Sekiguchi&Shibata05}, 
\citet{Ott11}, and \citet{CerdaDuran13} were among the first to study the BH formation of
 very rapidly rotating stars in GR but with simplified microphysics setups 
(e.g., polytropic equation of state (EOS) and/or a neutrino leakage scheme).
 In the context of multi-D simulations with multi-energy neutrino transport, 
 \cite{Chan&Muller18} were the first to report a
BH-forming 3D-GR simulation of a 40 solar mass zero-metallicity star.
They reported that the shock revival occurs at 0.35 s after bounce which is $\sim0.2$ s before the BH formation.
 More recently, \cite{Pan17} reported a similar result from post-Newtonian simulations
 of a 40 solar mass solar metallicity star, using a two-flavor IDSA scheme \citep{Liebendorfer09} and a leakage scheme for the heavy-flavor neutrinos.
\cite{Ott18} performed a series of 3D-GR simulations of solar-metallicity 
stars with multi-energy neutrino transport.
Their most massive progenitor (40 $M_{\odot}$) leads to the most energetic explosion among the 
computed models, where the BH formation was not followed due to the shorter simulation 
timescale ($T_{\rm pb}\sim 300$ ms postbounce). Comparing their 1D (non-exploding) 40 $M_{\odot}$ 
model, note that the BH formation would occur later
 than $T_{\rm pb}\sim 600 - 700$ ms.
 Thus far the BH formation has not yet been explored in
 the context of 3D simulations with two-moment neutrino 
transport schemes in GR as proposed in \citet{Shibata11,cardall13}.


In this Letter, we study the final fate of a 70 solar mass zero metallicity star 
 by performing 3D-GR core-collapse simulation with the
best available neutrino transport scheme \citep{KurodaT16}. Our results show that the BH 
formation time ($T_{\rm pb}\sim$ 300 ms) is significantly shorter than those 
 obtained in the previous multi-D simulations ($T_{\rm pb}\sim 1$ s)
 where the $40 M_{\odot}$ stars have only been investigated so far 
\citep{Chan&Muller18,Pan17}.
 Our results present the first numerical evidence to 
 extend the horizon of the validity of 
the fallback scenario up to a $70 M_{\odot}$ star 
where the neutrino-driven shock revival precedes the BH formation.

\section{Numerical Methods and Initial Model}
\label{sec2}
The numerical schemes in this work are essentially the same as those 
in \citet{KurodaT16}. 
Regarding the metric evolution, 
we evolve the standard BSSN variables
 \citep{Shibata95,Baumgarte99,Marronetti08} 
with a finite-difference scheme in space and with 
 a Runge-Kutta method in time, both in fourth-order accuracy. 
The gauge is specified by the ``1+log'' lapse and by the Gamma-driver-shift
condition.
Regarding the radiation-hydrodynamic evolution, the conservation equation 
$\nabla_\alpha T^{\alpha\beta}_{\rm (total)}=0$ is solved using
 the piecewise parabolic method \citep{Colella84,Hawke05}.
 $T^{\alpha\beta}_{\rm (total)}$ is the total stress-energy tensor, 
\begin{equation} T_{\rm (total)}^{\alpha\beta} = 
T_{\rm (fluid)}^{\alpha\beta} +\int d\varepsilon \sum_{\nu\in\nu_e,\bar\nu_e,\nu_x}T_{(\nu,\varepsilon)}^{\alpha\beta},
\label{TotalSETensor}
\end{equation}
where $T_{\rm (fluid)}^{\alpha\beta}$ and $T_{(\nu,\varepsilon)}^{\alpha\beta}$  
 are the stress-energy tensor of the fluid and
the neutrino radiation field, respectively.
We consider three-flavor of neutrinos ($\nu \in \nu_e,\bar\nu_e,\nu_x$) with 
$\nu_x$ denoting heavy-lepton neutrinos (i.e., $\nu_{\mu}, \nu_{\tau}$ and their anti-particles). $\varepsilon$ represents the neutrino energy measured in the 
comoving frame which logarithmically covers from 1 to 300 MeV with 12 energy bins.
Employing an M1 analytical closure scheme \citep{Shibata11},
 we solve spectral neutrino transport of the radiation energy and momentum,
based on the truncated moment formalism (e.g., \citet{KurodaT16,Roberts16,Ott18}).
We include the gravitational red- and Doppler-shift terms to follow the neutrino radiation field in highly curved spacetime around BH.
Regarding neutrino opacities, the 
standard weak interaction set in \citet{Bruenn85} plus nucleon-nucleon
 Bremsstrahlung \citep{Hannestad98} is taken into account.

We use a 70 solar mass zero-metallicity star 
 of \cite{Takahashi14}, which we refer to as Z70 below.
At the precollapse phase of Z70, the mass of the central iron core 
is $\sim2.8 M_{\odot}$ and the enclosed mass up to the helium layer
 is $\sim 31 M_{\odot}$.
By comparing with an ultra metal-poor star with a similar progenitor 
mass (a 75$M_{\odot}$ star of \citet{WHW02}, hereafter referred to as U75), the two progenitors are rather similar especially
 in the vicinity of the central cores, such as the progenitor's core-compact parameter  \citep{O'Connor11} 
($\xi_{2.5} \sim$ 1.00 and 0.86 for Z70 and U75, respectively), the central density and 
temperature (albeit being slightly higher ($\sim10\%$) for Z70).
We thus consider that our results would not change significantly even if we use 
U75.
For comparing with previous results, we also compute a 40 solar mass solar metallicity 
star of \citet{WHW02} that we refer to as S40.
Note that the compactness parameter
 of S40 is much smaller ($\xi_{2.5}\sim0.26$) than that of Z70.

We use the equation of state (EOS) by \citet{LSEOS} with a bulk incompressibility
modulus of $K$ = 220 MeV (LS220). 
The 3D computational domain is a cubic box
with 15,000 km width, and nested boxes with nine refinement
levels are embedded in the Cartesian coordinates. Each box contains $64^3$ cells and the
minimum grid size near the origin is $\Delta x=458$m.
The PNS core surface ($\sim10$ km) and stalled shock ($\sim110$-220 km)
are resolved by $\Delta x=458$m and $7.3$ km, respectively.


\section{Results}
\label{Sec:Black hole formation}
\begin{figure}
\begin{center}
\includegraphics[width=100mm,angle=-90.]{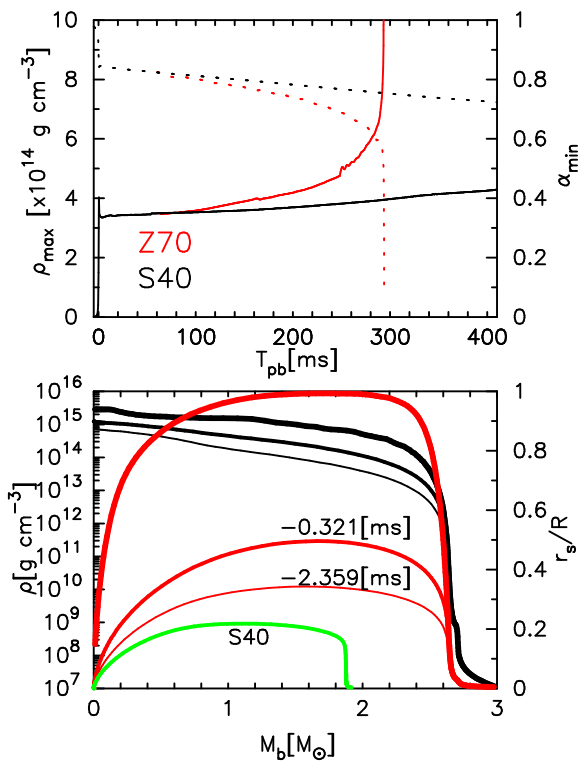}
\caption{{\it Top:} Time evolution of the maximum rest mass density (solid line) and minimum lapse function (dotted line) for Z70 (red line) and 
 S40 (black line), respectively.
{\it Bottom:} Spatial profiles of the rest mass density (black line) 
and the ratio $r_{\rm s}/R$ (red lines, see text) for Z70 as a function of the enclosed baryon mass for three representative time slices near the final simulation time.  ``-0.321'' and ``-2.359'' ms denote the 
time before the final simulation time 
($T_{\rm fin} = 293$ ms after bounce). The green line corresponds 
to $r_{\rm s}/R$ for S40 at $T_{\rm pb} = 400$ ms.
\label{f1}}
\end{center}
\end{figure}

The top panel of Fig.\ref{f1} shows the time evolution of the maximum 
(rest-mass) density $\rho_{\rm max}$ (sold lines) and the minimum lapse 
$\alpha_{\rm min}$ (dotted lines) for Z70 (red line) and 
S40 (black line), respectively. The maximum density at bounce 
($\rho_{\rm max}\sim 4 \times 10^{14}$ g cm$^{-3}$) is quite similar for the two models. After bounce, one can see that the increase of 
the maximum density of Z70 (red solid line) is significantly faster 
than that of S40 (black solid line). In Z70, 
 we stop the computation at $T_{\rm fin} = 293$ ms after bounce 
because it exceeds the (temperature) range of the EOS table used 
in our simulation.
In both Z70 and S40, the minimum lapse 
(dotted lines) shows a gradual decrease after bounce. At around 
$T_{\rm fin}$, it shows a drastic drop to $\alpha_{\rm min}=0.0645$ 
for Z70.

In the bottom panel of Fig.\ref{f1}, we explain
 how $T_{\rm fin}$ is related to the 
BH formation time. We show the profiles of 
 (angle-averaged) density (black lines) and a diagnostics to
 measure the BH formation as 
a function of the enclosed baryon mass $M_{\rm b}$ at some 
representative snapshots near $T_{\rm fin}$
 for Z70 (red lines) and at $T_{\rm pb} = 400$ ms 
for S40 (green line). We estimate the diagnostics by
 the ratio $r_{\rm s}/R$ where $r_{\rm s}$ and $R$ are the 
Schwarzschild radius and the radial coordinate, respectively
One can see that the maximum $r_{\rm s}/R$ 
is $\sim0.3$ at 2.359 ms before $T_{\rm fin}$ (the thin red line labelled 
 by ``-2.359 [ms]'') and rapidly increases with time, approaching to unity (precisely,
 $0.9932$) at $T_{\rm fin}$ (thickest red line), which we judge as 
 the epoch of the BH formation in this work. 
For the unambiguous definition of the BH formation, one 
 requires the implementation of the so-called apparent horizon 
 finder (e.g., \citet{thorn04}) in numerical relativity simulation, which we leave for future work.

At the (fiducial) BH formation time, the mass and the radius are 
$M_{\rm b(g),BH}\sim2.60(2.51)$ $M_\odot$ and 
$R_{\rm iso}\sim4$ km, respectively (see the thickest red line in
 the bottom panel of Fig.\ref{f1}).
By contrast, S40 shows a significantly less compact structure (green line)
 at the final simulation time ($T_{\rm pb} = 400 $ ms).
 The BH formation should occur much later, 
possibly when the mass shell at $R(M_{\rm b}=2.6 M_\odot)\sim10^9$ cm 
accretes onto the stalled shock. 
Using the same EOS (LS220), this expectation is in line 
with \citet{Chan&Muller18} who reported the BH formation at $T_{\rm pb}\sim 1$ s.

\begin{figure}
\begin{center}
\includegraphics[width=60mm,angle=-90.]{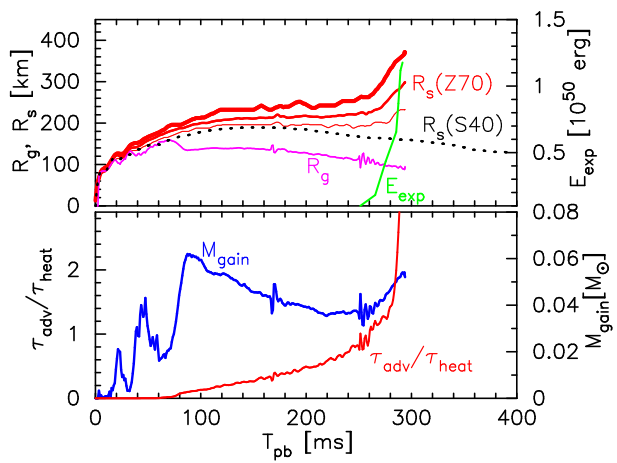}
\caption{{\it Top:} Postbounce evolution of the (angle-averaged) shock radii $R_{\rm s}$ (red, black), the gain radius $R_{\rm g}$ (magenta), and diagnostic explosion energy $E_{\rm exp}$ (green).
  The red lines correspond to the maximum, average, and minimum radii 
for Z70 and the black dotted line is (only) the average radius for S40.
 {\it Bottom:} The ratio of the advection timescale 
to the neutrino heating timescale in the gain region
 $\tau_{\rm adv}/\tau_{\rm heat}$ (red line) 
and the mass in the gain region $M_{\rm gain}$ (blue line) for Z70.
\label{f2}}
\end{center}
\end{figure}

 Fig. \ref{f2} displays the time evolution of the (angle-averaged) shock radius $R_{\rm s}$, the gain radius $R_{\rm g}$,
 the diagnostic explosion energy $E_{\rm exp}$ (see Eq.(2) of \cite{BMuller12a}), 
the ratio of the advection timescale to the neutrino-heating timescale
 in the gain region $\tau_{\rm adv}/\tau_{\rm heat}$ (e.g., \citet{Buras06b,KurodaT12}), and the mass in the gain region $M_{\rm gain}$, respectively.
For Z70 (red solid lines), one can clearly see
 the shock revival after $T_{\rm pb}\, \ga\, 260$ ms and rapid increase in $E_{\rm exp}$.
$E_{\rm exp}$ finally reaches $\sim1.2\times10^{50}$ erg which is, however, two orders of magnitude smaller than
the binding energy ahead of the shock.
 In this second collapse phase, the high neutrino emission makes the heating timescale shorter
 than the competing advection timescale in the gain region. 
Aided by strong convection behind the shock, the stalled shock is revived
  at $T_{\rm pb}\,\ga\,260$ ms
 ($\tau_{\rm adv}/\tau_{\rm heat} \geq 1$, red line in the bottom 
 panel of Fig.\ref{f2}). This also results in the increase 
in the gain mass (see the blue line) due to the shock expansion.
We note that the shock stall radii $R\sim200(180)$ km in Z70(S40) are rather larger than previous reports,
e.g., $\sim160$ km for S40 in \cite{Ott18}.
We consider that this comes most likely from a low numerical resolution of $\Delta X=7.3$ km around the shock surface 
(approximately two or more times coarser than that employed in \cite{Ott18}), 
 which could seed stronger prompt convection.

\begin{figure}
  \begin{center}
          \includegraphics[clip,width=85mm]{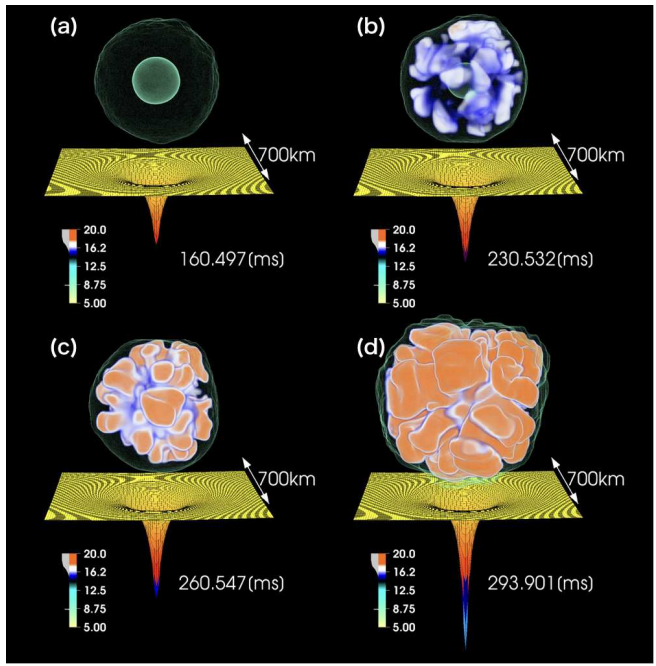}
          \caption{Snapshots of the entropy distribution 
($k_{\rm B}$ baryon$^{-1}$) for Z70 at $T_{\rm pb} \sim 160, 231, 261, $ and 294 ms (from {\it top left} to {\it bottom right}), respectively.
The sheet represents the lapse function ($\alpha$) on the $z=0$ plane.
  \label{f3}}
  \end{center}
\end{figure}

 Fig.\ref{f3} visualizes the hydrodynamic evolution in 3D after the shock stall 
(panel (a)), through the onset of convection (panels (b) and (c)), 
toward the shock-revival and the 
BH formation (panel (d)).
During the first $\sim160$ ms after bounce, the neutrino heating is still weak 
and the high entropy bubbles do not appear (panel (a)).
After $T_{\rm pb}\ga230$ ms, high entropy plumes with $s\ga15k_{\rm B}$ baryon$^{-1}$ are visible (panel (b)).
At this time, the mass in the gain region $M_{\rm gain}$ also starts to 
increase (Fig. \ref{f2}). Comparing panel (c) with (d), the expansion of the 
(merging) high entropy plumes is clearly seen, leading to the shock revival.
 The lapse function shows the steepest drop in the center 
(panel (d), see the cusp in the yellowish plane), which corresponds to 
 the BH formation. By expanding the shock radius in spherical harmonics,
 we find that the deviation of the shock from spherical symmetry 
(in the low-modes $\ell = 1,2$) is less than $\sim2$ $\%$. 
This clearly indicates that neutrino-driven convection dominates over the 
 SASI (standing accretion shock instability, \citet{Blondin03}) in this case.

\begin{figure}
  \begin{center}
          \includegraphics[clip,width=60mm,angle=-90.]{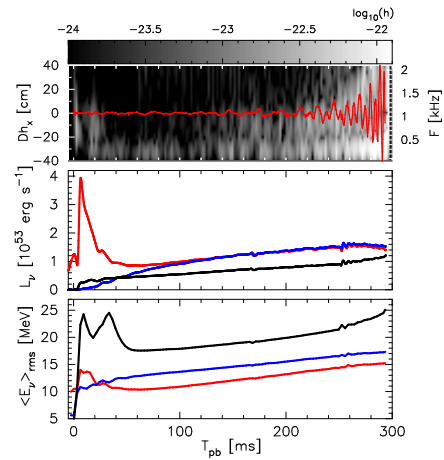}
          \caption{Gravitational waveform and its spectrogram(top),
            Neutrino luminosities(middle), and rms neutrino energies (bottom) 
 as a function of postbounce time for Z70.  Note in the 
top panel that $h_{\times}$ and $D$ denotes the GW amplitude of the cross polarization
 and the distance to the source, respectively.
            Red, blue, and black lines represent $\nu_e$, $\bar\nu_e$,
            and $\nu_x$, respectively.
  \label{f4}}
  \end{center}
\end{figure}

Finally, we show in Fig. \ref{f4}
the gravitational waveform for the cross mode and
the spectrogram of wave strain with assuming a source distance of $D=10$ kpc (top), 
the neutrino luminosity (middle), and rms neutrino energy (bottom) for Z70.
The waveform is extracted along the positive $z$-axis via a standard quadrupole formula
 \citep{KurodaT14} where the GR corrections are taken into account (e.g., \cite{Shibata&Sekiguchi03}).
The rms neutrino energies and neutrino luminosities
are obtained from the radiation energy and flux 
in the comoving frame at $R_{\rm iso}=400$ km, respectively.

The GW amplitude stays at a small value $Dh\la10$ cm before the shock expansion occurs.
The strong GW emission most likely originates from the $g$-mode oscillation 
of the PNS and strong convection motion behind the shock (e.g., panels (c) and (d) in Fig.\ref{f3}).
This can be understood from broad band emissions $100\la F\la3000$ Hz in the spectrogram.
Such a broad-band feature is consistent with \cite{Ott11}
who showed the GW amplitudes reaching $Dh\sim$30 cm for their non-rotating model.
Note that there are significant differences between our numerical method and the ones 
in \cite{Ott11} who assumed octant symmetry with different numerical resolution than ours, used polytropic EOS, and omitted neutrino radiation.

The neutrino luminosity and rms energy of electron type (anti-)neutrinos show a decreasing
and plateau trend for $T_{\rm pb}\ga260$ ms, respectively.
On the other hand, heavy-lepton neutrinos show a rapid increase both in
 the luminosity and rms energy\footnote{
We find a transient increment at $T_{\rm pb}\sim40$ ms in $\langle E_{\nu_x}\rangle_{\rm rms}$.
We consider that this might be due to stronger shock and
prompt convection motions which upscatter neutrinos.
Since this transient increment diminishes for the mean energy $\langle E_{\nu_x}\rangle_{\rm mean}$,
only a small portion of neutrinos are affected.
It could be the artifact of mesh refinement boundary and/or
relatively low numerical resolution.  Simulation with large number of
grids is necessary to clarify this issue.}.
As previously identified in 1D full-GR simulations with Boltzmann neutrino 
transport \citep{Liebendorfer04}, these features are commonly observed
 in the literature, due to rapid contraction of the PNS to the forming BH
(see also, \citet{Sumiyoshi07,Fischer09,Hempel12}). The detection
 of the short-live ($\sim 300$ ms after bounce) neutrino signals are basically 
 limited to Galactic events (see \citet{mirizzi16} for a review). However,
 further study would be needed to clarify the contribution of these
BH forming massive stars to the prediction of the diffuse supernova neutrino 
background (e.g., \citet{lunardini09,Horiuchi18}). 

\section{Summary and Discussions}
\label{Sec:Summary and Discussion}
 We have presented the results of full 3D-GR core-collapse simulations of 
 massive stars with three-flavor spectral neutrino transport using the M1 closure scheme.
 Employing a 70 solar mass zero metallicity progenitor,
 we self-consistently followed the 3D hydrodynamics from 
the onset of gravitational core-collapse until the second collapse of 
the PNS, leading to the BH formation.
 We showed that the BH formation occurs at the post-bounce time of 
$T_{\rm pb}\sim 300$ ms for the 70 $M_{\odot}$ star, which is significantly 
 earlier than found in the literature. At a few $\sim10$ ms before the BH 
formation, the neutrino-driven shock revival was obtained, which is 
aided by violent convection behind the shock. 
Although it requires much longer simulation time to evaluate the final explosion energy $E_{\rm exp}$
and also determine if the mass ejection occurs,
$E_{\rm exp}$, at the BH formation time, is significantly lower than the binding energy ahead of the shock.
This indicates that the mass ejection would be hardly expected.
Our full 3D-GR core-collapse simulations, however, present the first evidence to 
 validate the BH formation by fallback \citep{Heger03} up to the $70 M_{\odot}$ star, 
where the neutrino-driven shock revival precedes the BH formation.
We also presented analysis on the neutrino emission which possesses characteristic signatures of the second collapse.

Note that this is the first simulation where the
 70 $M_\odot$ of \cite{Takahashi14} was employed.
We shortly compare our results with previous works that focused on the 
 BH formation.
\citet{O'Connor11} studied in their extensive 1D simulations the impact of progenitors and
 EOSs on the BH formation while simplifying neutrino physics by a leakage 
scheme. Their U75 model using the same EOS (LS220) 
shows $M_{\rm b(g),BH}\sim$2.592(2.498) $M_\odot$, which is close to that 
$M_{\rm b(g),BH}\sim2.60(2.51)$ $M_\odot$ for Z70 in this work.
Regarding S40, the BH formation does not occur in the first 300 ms after bounce,
 which is consistent with multi-D results by \cite{Chan&Muller18,Pan17}.

In this study, we have presented results of only the two progenitors.
 Although one of them leads to the BH formation with a 
mass of $\sim2.6$ $M_\odot$, it is a long way to understand the final 
 BH mass ($\sim8$ to $\sim30$ $M_\odot$) observed in the GW events
\citep{GW150914,GW151226,GW170104}.
 We need to follow a much longer evolution 
up to the order of $T_{\rm pb} \sim 10$ s until the entire helium core (at 
 a radius of  $10^9$ cm) accretes to the BH.
To do this, one needs to implement both the apparent horizon finder and some special treatments for neutrinos and fluids within the apparent horizon (e.g., \cite{Hawke05}).
 Understanding the dynamics and evolution of the BH-forming massive stars 
should progress not only with such an update 
in numerical techniques but also with the advance in 
 binary stellar evolutionary calculations (e.g., \citet{marchant16}).
\vspace{-0.5cm}
\section*{Acknowledgements}
This research was supported by the European Research Council (ERC; FP7) under ERC Advanced Grant FISH-321263 (TK and FKT),
ERC Starting Grant EUROPIUM-677912 (TK),
JSPS KAKENHI Grant Number (JP15KK0173 (KK),
JP17H05206, JP17K14306, JP15H00789, JP15H01039 (TK), and
JP17H01130, JP17H06364 (KK and TT)),
and JICFuS as a priority issue to be tackled by using the Post `K' Computer.
Numerical computations were carried out on Cray XC30 at CfCA, NAOJ.



\bibliographystyle{mnras}
\bibliography{mybib} 

\begin{thebibliography}{}
\makeatletter
\relax
\def\mn@urlcharsother{\let\do\@makeother \do\$\do\&\do\#\do\^\do\_\do\%\do\~}
\def\mn@doi{\begingroup\mn@urlcharsother \@ifnextchar [ {\mn@doi@}
  {\mn@doi@[]}}
\def\mn@doi@[#1]#2{\def\@tempa{#1}\ifx\@tempa\@empty \href
  {http://dx.doi.org/#2} {doi:#2}\else \href {http://dx.doi.org/#2} {#1}\fi
  \endgroup}
\def\mn@eprint#1#2{\mn@eprint@#1:#2::\@nil}
\def\mn@eprint@arXiv#1{\href {http://arxiv.org/abs/#1} {{\tt arXiv:#1}}}
\def\mn@eprint@dblp#1{\href {http://dblp.uni-trier.de/rec/bibtex/#1.xml}
  {dblp:#1}}
\def\mn@eprint@#1:#2:#3:#4\@nil{\def\@tempa {#1}\def\@tempb {#2}\def\@tempc
  {#3}\ifx \@tempc \@empty \let \@tempc \@tempb \let \@tempb \@tempa \fi \ifx
  \@tempb \@empty \def\@tempb {arXiv}\fi \@ifundefined
  {mn@eprint@\@tempb}{\@tempb:\@tempc}{\expandafter \expandafter \csname
  mn@eprint@\@tempb\endcsname \expandafter{\@tempc}}}

\bibitem[\protect\citeauthoryear{{Abbott} et~al.,}{{Abbott}
  et~al.}{2016a}]{GW150914}
{Abbott} B.~P.,  et~al., 2016a, \mn@doi [Physical Review Letters]
  {10.1103/PhysRevLett.116.061102}, \href
  {http://ads.nao.ac.jp/abs/2016PhRvL.116f1102A} {116, 061102}

\bibitem[\protect\citeauthoryear{{Abbott} et~al.,}{{Abbott}
  et~al.}{2016b}]{GW151226}
{Abbott} B.~P.,  et~al., 2016b, \mn@doi [Physical Review Letters]
  {10.1103/PhysRevLett.116.241103}, \href
  {http://cdsads.u-strasbg.fr/abs/2016PhRvL.116x1103A} {116, 241103}

\bibitem[\protect\citeauthoryear{{Abbott} et~al.,}{{Abbott}
  et~al.}{2016c}]{abbott16}
{Abbott} B.~P.,  et~al., 2016c, \mn@doi [\apjl] {10.3847/2041-8205/818/2/L22},
  \href {http://ads.nao.ac.jp/abs/2016ApJ...818L..22A} {818, L22}

\bibitem[\protect\citeauthoryear{{Abbott} et~al.,}{{Abbott}
  et~al.}{2017}]{GW170104}
{Abbott} B.~P.,  et~al., 2017, \mn@doi [Physical Review Letters]
  {10.1103/PhysRevLett.118.221101}, \href
  {http://cdsads.u-strasbg.fr/abs/2017PhRvL.118v1101A} {118, 221101}

\bibitem[\protect\citeauthoryear{{Baumgarte} \& {Shapiro}}{{Baumgarte} \&
  {Shapiro}}{1999}]{Baumgarte99}
{Baumgarte} T.~W.,  {Shapiro} S.~L.,  1999, \mn@doi [\prd]
  {10.1103/PhysRevD.59.024007}, \href
  {http://ads.nao.ac.jp/abs/1999PhRvD..59b4007B} {59, 024007}

\bibitem[\protect\citeauthoryear{{Belczynski}, {Dominik}, {Bulik},
  {O'Shaughnessy}, {Fryer}  \& {Holz}}{{Belczynski} et~al.}{2010}]{bel10}
{Belczynski} K.,  {Dominik} M.,  {Bulik} T.,  {O'Shaughnessy} R.,  {Fryer} C.,
   {Holz} D.~E.,  2010, \mn@doi [\apjl] {10.1088/2041-8205/715/2/L138}, \href
  {http://ads.nao.ac.jp/abs/2010ApJ...715L.138B} {715, L138}

\bibitem[\protect\citeauthoryear{{Belczynski}, {Buonanno}, {Cantiello},
  {Fryer}, {Holz}, {Mandel}, {Miller}  \& {Walczak}}{{Belczynski}
  et~al.}{2014}]{bel14}
{Belczynski} K.,  {Buonanno} A.,  {Cantiello} M.,  {Fryer} C.~L.,  {Holz}
  D.~E.,  {Mandel} I.,  {Miller} M.~C.,   {Walczak} M.,  2014, \mn@doi [\apj]
  {10.1088/0004-637X/789/2/120}, \href
  {http://ads.nao.ac.jp/abs/2014ApJ...789..120B} {789, 120}

\bibitem[\protect\citeauthoryear{{Blondin}, {Mezzacappa}  \&
  {DeMarino}}{{Blondin} et~al.}{2003}]{Blondin03}
{Blondin} J.~M.,  {Mezzacappa} A.,   {DeMarino} C.,  2003, \mn@doi [\apj]
  {10.1086/345812}, \href {http://ads.nao.ac.jp/abs/2003ApJ...584..971B} {584,
  971}

\bibitem[\protect\citeauthoryear{{Bruenn}}{{Bruenn}}{1985}]{Bruenn85}
{Bruenn} S.~W.,  1985, \mn@doi [\apjs] {10.1086/191056}, \href
  {http://cdsads.u-strasbg.fr/abs/1985ApJS...58..771B} {58, 771}

\bibitem[\protect\citeauthoryear{{Buras}, {Janka}, {Rampp}  \&
  {Kifonidis}}{{Buras} et~al.}{2006}]{Buras06b}
{Buras} R.,  {Janka} H.-T.,  {Rampp} M.,   {Kifonidis} K.,  2006, \mn@doi
  [\aap] {10.1051/0004-6361:20054654}, \href
  {http://ads.nao.ac.jp/abs/2006A%26A...457..281B} {457, 281}

\bibitem[\protect\citeauthoryear{{Cardall}, {Endeve}  \&
  {Mezzacappa}}{{Cardall} et~al.}{2013}]{cardall13}
{Cardall} C.~Y.,  {Endeve} E.,   {Mezzacappa} A.,  2013, \mn@doi [\prd]
  {10.1103/PhysRevD.88.023011}, \href
  {http://adsabs.harvard.edu/abs/2013PhRvD..88b3011C} {88, 023011}

\bibitem[\protect\citeauthoryear{{Cerd{\'a}-Dur{\'a}n}, {DeBrye}, {Aloy},
  {Font}  \& {Obergaulinger}}{{Cerd{\'a}-Dur{\'a}n}
  et~al.}{2013}]{CerdaDuran13}
{Cerd{\'a}-Dur{\'a}n} P.,  {DeBrye} N.,  {Aloy} M.~A.,  {Font} J.~A.,
  {Obergaulinger} M.,  2013, \mn@doi [\apjl] {10.1088/2041-8205/779/2/L18},
  \href {http://cdsads.u-strasbg.fr/abs/2013ApJ...779L..18C} {779, L18}

\bibitem[\protect\citeauthoryear{{Chan}, {M{\"u}ller}, {Heger}, {Pakmor}  \&
  {Springel}}{{Chan} et~al.}{2018}]{Chan&Muller18}
{Chan} C.,  {M{\"u}ller} B.,  {Heger} A.,  {Pakmor} R.,   {Springel} V.,  2018,
  \mn@doi [\apjl] {10.3847/2041-8213/aaa28c}, \href
  {http://cdsads.u-strasbg.fr/abs/2018ApJ...852L..19C} {852, L19}

\bibitem[\protect\citeauthoryear{{Colella} \& {Woodward}}{{Colella} \&
  {Woodward}}{1984}]{Colella84}
{Colella} P.,  {Woodward} P.~R.,  1984, \mn@doi [Journal of Computational
  Physics] {10.1016/0021-9991(84)90143-8}, \href
  {http://cdsads.u-strasbg.fr/abs/1984JCoPh..54..174C} {54, 174}

\bibitem[\protect\citeauthoryear{{Fischer}, {Whitehouse}, {Mezzacappa},
  {Thielemann}  \& {Liebend{\"o}rfer}}{{Fischer} et~al.}{2009}]{Fischer09}
{Fischer} T.,  {Whitehouse} S.~C.,  {Mezzacappa} A.,  {Thielemann} F.-K.,
  {Liebend{\"o}rfer} M.,  2009, \mn@doi [\aap] {10.1051/0004-6361/200811055},
  \href {http://cdsads.u-strasbg.fr/abs/2009A%26A...499....1F} {499, 1}

\bibitem[\protect\citeauthoryear{{Fryer}, {Woosley}  \& {Hartmann}}{{Fryer}
  et~al.}{1999}]{Fryer99}
{Fryer} C.~L.,  {Woosley} S.~E.,   {Hartmann} D.~H.,  1999, \mn@doi [\apj]
  {10.1086/307992}, \href {http://cdsads.u-strasbg.fr/abs/1999ApJ...526..152F}
  {526, 152}

\bibitem[\protect\citeauthoryear{{Hannestad} \& {Raffelt}}{{Hannestad} \&
  {Raffelt}}{1998}]{Hannestad98}
{Hannestad} S.,  {Raffelt} G.,  1998, \mn@doi [\apj] {10.1086/306303}, \href
  {http://ads.nao.ac.jp/abs/1998ApJ...507..339H} {507, 339}

\bibitem[\protect\citeauthoryear{{Hawke}, {L{\"o}ffler}  \& {Nerozzi}}{{Hawke}
  et~al.}{2005}]{Hawke05}
{Hawke} I.,  {L{\"o}ffler} F.,   {Nerozzi} A.,  2005, \mn@doi [\prd]
  {10.1103/PhysRevD.71.104006}, \href
  {http://cdsads.u-strasbg.fr/abs/2005PhRvD..71j4006H} {71, 104006}

\bibitem[\protect\citeauthoryear{{Heger}, {Fryer}, {Woosley}, {Langer}  \&
  {Hartmann}}{{Heger} et~al.}{2003}]{Heger03}
{Heger} A.,  {Fryer} C.~L.,  {Woosley} S.~E.,  {Langer} N.,   {Hartmann} D.~H.,
   2003, \mn@doi [\apj] {10.1086/375341}, \href
  {http://cdsads.u-strasbg.fr/abs/2003ApJ...591..288H} {591, 288}

\bibitem[\protect\citeauthoryear{{Hempel}, {Fischer}, {Schaffner-Bielich}  \&
  {Liebend{\"o}rfer}}{{Hempel} et~al.}{2012}]{Hempel12}
{Hempel} M.,  {Fischer} T.,  {Schaffner-Bielich} J.,   {Liebend{\"o}rfer} M.,
  2012, \mn@doi [\apj] {10.1088/0004-637X/748/1/70}, \href
  {http://cdsads.u-strasbg.fr/abs/2012ApJ...748...70H} {748, 70}

\bibitem[\protect\citeauthoryear{{Horiuchi}, {Sumiyoshi}, {Nakamura},
  {Fischer}, {Summa}, {Takiwaki}, {Janka}  \& {Kotake}}{{Horiuchi}
  et~al.}{2018}]{Horiuchi18}
{Horiuchi} S.,  {Sumiyoshi} K.,  {Nakamura} K.,  {Fischer} T.,  {Summa} A.,
  {Takiwaki} T.,  {Janka} H.-T.,   {Kotake} K.,  2018, \mn@doi [\mnras]
  {10.1093/mnras/stx3271}, \href
  {http://cdsads.u-strasbg.fr/abs/2018MNRAS.475.1363H} {475, 1363}

\bibitem[\protect\citeauthoryear{{Janka}, {Melson}  \& {Summa}}{{Janka}
  et~al.}{2016}]{janka16}
{Janka} H.-T.,  {Melson} T.,   {Summa} A.,  2016, Annual Review of Nuclear and
  Particle Science, \href {http://adsabs.harvard.edu/abs/2016arXiv160205576J}
  {}

\bibitem[\protect\citeauthoryear{{Kinugawa}, {Nakano}  \&
  {Nakamura}}{{Kinugawa} et~al.}{2016}]{Kinugawa16}
{Kinugawa} T.,  {Nakano} H.,   {Nakamura} T.,  2016, \mn@doi [Progress of
  Theoretical and Experimental Physics] {10.1093/ptep/ptw143}, \href
  {http://cdsads.u-strasbg.fr/abs/2016PTEP.2016j3E01K} {2016, 103E01}

\bibitem[\protect\citeauthoryear{{Kotake}, {Sumiyoshi}, {Yamada}, {Takiwaki},
  {Kuroda}, {Suwa}  \& {Nagakura}}{{Kotake} et~al.}{2012}]{Kotake12_ptep}
{Kotake} K.,  {Sumiyoshi} K.,  {Yamada} S.,  {Takiwaki} T.,  {Kuroda} T.,
  {Suwa} Y.,   {Nagakura} H.,  2012, \mn@doi [Progress of Theoretical and
  Experimental Physics] {10.1093/ptep/pts009}, \href
  {http://ads.nao.ac.jp/abs/2012PTEP.2012aA301K} {2012, 010000}

\bibitem[\protect\citeauthoryear{{Kuroda}, {Kotake}  \& {Takiwaki}}{{Kuroda}
  et~al.}{2012}]{KurodaT12}
{Kuroda} T.,  {Kotake} K.,   {Takiwaki} T.,  2012, \mn@doi [\apj]
  {10.1088/0004-637X/755/1/11}, \href
  {http://ads.nao.ac.jp/abs/2012ApJ...755...11K} {755, 11}

\bibitem[\protect\citeauthoryear{{Kuroda}, {Takiwaki}  \& {Kotake}}{{Kuroda}
  et~al.}{2014}]{KurodaT14}
{Kuroda} T.,  {Takiwaki} T.,   {Kotake} K.,  2014, \mn@doi [\prd]
  {10.1103/PhysRevD.89.044011}, \href
  {http://ads.nao.ac.jp/abs/2014PhRvD..89d4011K} {89, 044011}

\bibitem[\protect\citeauthoryear{{Kuroda}, {Takiwaki}  \& {Kotake}}{{Kuroda}
  et~al.}{2016}]{KurodaT16}
{Kuroda} T.,  {Takiwaki} T.,   {Kotake} K.,  2016, \mn@doi [\apjs]
  {10.3847/0067-0049/222/2/20}, \href
  {http://cdsads.u-strasbg.fr/abs/2016ApJS..222...20K} {222, 20}

\bibitem[\protect\citeauthoryear{{Langer}}{{Langer}}{2012}]{Langer12}
{Langer} N.,  2012, \mn@doi [\araa] {10.1146/annurev-astro-081811-125534},
  \href {http://cdsads.u-strasbg.fr/abs/2012ARA%26A..50..107L} {50, 107}

\bibitem[\protect\citeauthoryear{{Lattimer} \& {Swesty}}{{Lattimer} \&
  {Swesty}}{1991}]{LSEOS}
{Lattimer} J.~M.,  {Swesty} F.,  1991, \mn@doi [Nuclear Physics A]
  {10.1016/0375-9474(91)90452-C}, \href
  {http://cdsads.u-strasbg.fr/abs/1991NuPhA.535..331L} {535, 331}

\bibitem[\protect\citeauthoryear{{Lentz} et~al.,}{{Lentz}
  et~al.}{2015}]{lentz15}
{Lentz} E.~J.,  et~al., 2015, \mn@doi [\apjl] {10.1088/2041-8205/807/2/L31},
  \href {http://ads.nao.ac.jp/abs/2015ApJ...807L..31L} {807, L31}

\bibitem[\protect\citeauthoryear{{Liebend{\"o}rfer}, {Messer}, {Mezzacappa},
  {Bruenn}, {Cardall}  \& {Thielemann}}{{Liebend{\"o}rfer}
  et~al.}{2004}]{Liebendorfer04}
{Liebend{\"o}rfer} M.,  {Messer} O.~E.~B.,  {Mezzacappa} A.,  {Bruenn} S.~W.,
  {Cardall} C.~Y.,   {Thielemann} F.-K.,  2004, \mn@doi [\apjs]
  {10.1086/380191}, \href {http://cdsads.u-strasbg.fr/abs/2004ApJS..150..263L}
  {150, 263}

\bibitem[\protect\citeauthoryear{{Liebend{\"o}rfer}, {Whitehouse}  \&
  {Fischer}}{{Liebend{\"o}rfer} et~al.}{2009}]{Liebendorfer09}
{Liebend{\"o}rfer} M.,  {Whitehouse} S.~C.,   {Fischer} T.,  2009, \mn@doi
  [\apj] {10.1088/0004-637X/698/2/1174}, \href
  {http://ads.nao.ac.jp/abs/2009ApJ...698.1174L} {698, 1174}

\bibitem[\protect\citeauthoryear{{Lunardini}}{{Lunardini}}{2009}]{lunardini09}
{Lunardini} C.,  2009, \mn@doi [Physical Review Letters]
  {10.1103/PhysRevLett.102.231101}, \href
  {http://ads.nao.ac.jp/abs/2009PhRvL.102w1101L} {102, 231101}

\bibitem[\protect\citeauthoryear{{Marchant}, {Langer}, {Podsiadlowski},
  {Tauris}  \& {Moriya}}{{Marchant} et~al.}{2016}]{marchant16}
{Marchant} P.,  {Langer} N.,  {Podsiadlowski} P.,  {Tauris} T.~M.,   {Moriya}
  T.~J.,  2016, \mn@doi [\aap] {10.1051/0004-6361/201628133}, \href
  {http://ads.nao.ac.jp/abs/2016A%26A...588A..50M} {588, A50}

\bibitem[\protect\citeauthoryear{{Marronetti}, {Tichy}, {Br{\"u}gmann},
  {Gonz{\'a}lez}  \& {Sperhake}}{{Marronetti} et~al.}{2008}]{Marronetti08}
{Marronetti} P.,  {Tichy} W.,  {Br{\"u}gmann} B.,  {Gonz{\'a}lez} J.,
  {Sperhake} U.,  2008, \mn@doi [\prd] {10.1103/PhysRevD.77.064010}, \href
  {http://cdsads.u-strasbg.fr/abs/2008PhRvD..77f4010M} {77, 064010}

\bibitem[\protect\citeauthoryear{{Mirizzi}, {Tamborra}, {Janka}, {Saviano},
  {Scholberg}, {Bollig}, {H{\"u}depohl}  \& {Chakraborty}}{{Mirizzi}
  et~al.}{2016}]{mirizzi16}
{Mirizzi} A.,  {Tamborra} I.,  {Janka} H.-T.,  {Saviano} N.,  {Scholberg} K.,
  {Bollig} R.,  {H{\"u}depohl} L.,   {Chakraborty} S.,  2016, \mn@doi [Nuovo
  Cimento Rivista Serie] {10.1393/ncr/i2016-10120-8}, \href
  {http://ads.nao.ac.jp/abs/2016NCimR..39....1M} {39, 1}

\bibitem[\protect\citeauthoryear{{M{\"u}ller}, {Janka}  \&
  {Marek}}{{M{\"u}ller} et~al.}{2012}]{BMuller12a}
{M{\"u}ller} B.,  {Janka} H.-T.,   {Marek} A.,  2012, \mn@doi [\apj]
  {10.1088/0004-637X/756/1/84}, \href
  {http://adsabs.harvard.edu/abs/2012ApJ...756...84M} {756, 84}

\bibitem[\protect\citeauthoryear{{O'Connor} \& {Ott}}{{O'Connor} \&
  {Ott}}{2011}]{O'Connor11}
{O'Connor} E.,  {Ott} C.~D.,  2011, \mn@doi [\apj]
  {10.1088/0004-637X/730/2/70}, \href
  {http://ads.nao.ac.jp/abs/2011ApJ...730...70O} {730, 70}

\bibitem[\protect\citeauthoryear{{Ott} et~al.,}{{Ott} et~al.}{2011}]{Ott11}
{Ott} C.~D.,  et~al., 2011, \mn@doi [Physical Review Letters]
  {10.1103/PhysRevLett.106.161103}, \href
  {http://cdsads.u-strasbg.fr/abs/2011PhRvL.106p1103O} {106, 161103}

\bibitem[\protect\citeauthoryear{{Ott}, {Roberts}, {da Silva Schneider},
  {Fedrow}, {Haas}  \& {Schnetter}}{{Ott} et~al.}{2018}]{Ott18}
{Ott} C.~D.,  {Roberts} L.~F.,  {da Silva Schneider} A.,  {Fedrow} J.~M.,
  {Haas} R.,   {Schnetter} E.,  2018, \mn@doi [\apjl]
  {10.3847/2041-8213/aaa967}, \href
  {http://cdsads.u-strasbg.fr/abs/2018ApJ...855L...3O} {855, L3}

\bibitem[\protect\citeauthoryear{{Pan}, {Liebend{\"o}rfer}, {Couch}  \&
  {Thielemann}}{{Pan} et~al.}{2017}]{Pan17}
{Pan} K.-C.,  {Liebend{\"o}rfer} M.,  {Couch} S.~M.,   {Thielemann} F.-K.,
  2017, preprint, \href {http://cdsads.u-strasbg.fr/abs/2017arXiv171001690P} {}
  (\mn@eprint {arXiv} {1710.01690})

\bibitem[\protect\citeauthoryear{{Roberts}, {Ott}, {Haas}, {O'Connor}, {Diener}
   \& {Schnetter}}{{Roberts} et~al.}{2016}]{Roberts16}
{Roberts} L.~F.,  {Ott} C.~D.,  {Haas} R.,  {O'Connor} E.~P.,  {Diener} P.,
  {Schnetter} E.,  2016, \mn@doi [\apj] {10.3847/0004-637X/831/1/98}, \href
  {http://ads.nao.ac.jp/abs/2016ApJ...831...98R} {831, 98}

\bibitem[\protect\citeauthoryear{{Sekiguchi} \& {Shibata}}{{Sekiguchi} \&
  {Shibata}}{2005}]{Sekiguchi&Shibata05}
{Sekiguchi} Y.-I.,  {Shibata} M.,  2005, \mn@doi [\prd]
  {10.1103/PhysRevD.71.084013}, \href
  {http://cdsads.u-strasbg.fr/abs/2005PhRvD..71h4013S} {71, 084013}

\bibitem[\protect\citeauthoryear{{Shibata} \& {Nakamura}}{{Shibata} \&
  {Nakamura}}{1995}]{Shibata95}
{Shibata} M.,  {Nakamura} T.,  1995, \mn@doi [\prd] {10.1103/PhysRevD.52.5428},
  \href {http://ads.nao.ac.jp/abs/1995PhRvD..52.5428S} {52, 5428}

\bibitem[\protect\citeauthoryear{{Shibata} \& {Sekiguchi}}{{Shibata} \&
  {Sekiguchi}}{2003}]{Shibata&Sekiguchi03}
{Shibata} M.,  {Sekiguchi} Y.-I.,  2003, \mn@doi [\prd]
  {10.1103/PhysRevD.68.104020}, \href
  {http://cdsads.u-strasbg.fr/abs/2003PhRvD..68j4020S} {68, 104020}

\bibitem[\protect\citeauthoryear{{Shibata}, {Kiuchi}, {Sekiguchi}  \&
  {Suwa}}{{Shibata} et~al.}{2011}]{Shibata11}
{Shibata} M.,  {Kiuchi} K.,  {Sekiguchi} Y.,   {Suwa} Y.,  2011, Progress of
  Theoretical Physics, \href {http://ads.nao.ac.jp/abs/2011PThPh.125.1255S}
  {125, 1255}

\bibitem[\protect\citeauthoryear{{Sumiyoshi}, {Yamada}  \&
  {Suzuki}}{{Sumiyoshi} et~al.}{2007}]{Sumiyoshi07}
{Sumiyoshi} K.,  {Yamada} S.,   {Suzuki} H.,  2007, \mn@doi [\apj]
  {10.1086/520876}, \href {http://cdsads.u-strasbg.fr/abs/2007ApJ...667..382S}
  {667, 382}

\bibitem[\protect\citeauthoryear{{Takahashi}, {Umeda}  \&
  {Yoshida}}{{Takahashi} et~al.}{2014}]{Takahashi14}
{Takahashi} K.,  {Umeda} H.,   {Yoshida} T.,  2014, \mn@doi [\apj]
  {10.1088/0004-637X/794/1/40}, \href
  {http://cdsads.u-strasbg.fr/abs/2014ApJ...794...40T} {794, 40}

\bibitem[\protect\citeauthoryear{{Thornburg}}{{Thornburg}}{2004}]{thorn04}
{Thornburg} J.,  2004, \mn@doi [Classical and Quantum Gravity]
  {10.1088/0264-9381/21/2/026}, \href
  {http://adsabs.harvard.edu/abs/2004CQGra..21..743T} {21, 743}

\bibitem[\protect\citeauthoryear{{Woosley}, {Heger}  \& {Weaver}}{{Woosley}
  et~al.}{2002}]{WHW02}
{Woosley} S.~E.,  {Heger} A.,   {Weaver} T.~A.,  2002, \mn@doi [Reviews of
  Modern Physics] {10.1103/RevModPhys.74.1015}, \href
  {http://cdsads.u-strasbg.fr/abs/2002RvMP...74.1015W} {74, 1015}

\makeatother
\end{thebibliography}




\bsp	
\label{lastpage}
\end{document}